\documentclass[aps,pra,preprint,showkeys,superscriptaddress]{revtex4-2}
\usepackage{amsmath}
\usepackage{graphicx}

\begin{document}


\title{Revisiting optical rotation in helically-coiled fibers}


\author{Chun-Fang Li}
\email[Corresponding author. E-mail address: ]{cfli@shu.edu.cn}
\affiliation{Department of Physics, Shanghai University, 99 Shangda Road, 200444 Shanghai, China}

\author{Zhi-Juan Hu}
\affiliation{Department of Physics, Shanghai Normal University, 100 Guilin Road, 200233 Shanghai, China}



\begin{abstract}

The interpretation of optical rotation in optically active media as circular birefringence has persisted for over two centuries, yet the inherent fallacy in this phenomenological theory remains unnoticed. Recently, we employed logical reasoning to demonstrate that isotropic chiral media, a kind of optically active media, do not exhibit circular birefringence. This finding implies that the Jones vector is not able to completely describe the polarization state of a plane light wave. To further explore the reason, here we revisit the phenomenon of optical rotation in helically-coiled optical fibers. 

Firstly, we use similar logical reasoning to prove that helically coiled fibers do not exhibit circular birefringence, either. Secondly, based on the experimental observations of Papp and Harms, we argue that the Jones vector is mathematically an entity in the local reference frame associated with the propagation direction. It cannot completely describe the state of polarization relative to the laboratory reference frame. Meanwhile, we also demonstrate that the rotation observed by Papp and Harms reflects the rotation of the Tang frame relative to the Serret-Frenet frame.

\end{abstract}



\maketitle


\newpage

\section{Introduction}

Optical rotation is one of the most fundamental optical phenomena in nature. It refers to the rotation of the polarization plane of linearly-polarized light as it travels through optically active media \cite{Hecht, Gold}. Chiral substances \cite{Barr, Lind-STV, Lekn} and some kinds of crystals \cite{Kami, Kons-KNE, Chou-KH} are such media.
Circular birefringence, proposed by Fresnel \cite{Fres}, is commonly believed \cite{Hecht, Gold, Lekn, Barr, Kami, Kons-KNE, Chou-KH, Lind-STV, Ghos-F, Ren-LLL, Beut-SWRF} to be the mechanism for the optical rotation. 
Because a linearly polarized wave can be represented as a superposition of right-handed and left-handed circularly polarized states, Fresnel suggested in 1822 that these two forms of circularly polarized light propagate at different velocities.
Despite the publication of a huge number of research papers on optical rotation since then, no one seems to have spotted that the notion of circular birefringence contains a logical fallacy similar to that in Aristotle's theory of free-fall \cite{Gali, Skaf}. 
Recently we proved \cite{Hu-L, Hu-L24}, through logical reasoning analogous to that of Galileo \cite{Gali} in refuting Aristotle's theory, that there is no circular birefringence in a chiral medium. The two orthogonal circularly polarized waves propagate at the same velocity. This result does not contradict the phenomenon of circular double refraction that was observed \cite{Fres, Ghos-F} at the surface of the chiral medium. The key points are as follows.

Though the optical rotation in a chiral medium is commonly formulated as the rotation of the polarization plane of linearly-polarized light, the orientation of the polarization ellipse of any elliptically-polarized light is also rotated \cite{Beut-SWRF}. Hence, as a special case of elliptically polarized light, the state of polarization of circularly polarized light is rotated as well. Unfortunately, this kind of rotation has been completely omitted from the conventional interpretation. 
The peculiarity of such a rotation is that it gives rise to a phase shift in proportion to the propagation distance \cite{Hu-L}.
More importantly, after traveling the same distance in one particular chiral medium, the two orthogonal circularly polarized waves acquire, from the rotation of their polarization states, phase shifts that are equal in magnitude but opposite in sign. This is why they are mistakenly believed to propagate at different velocities under the assumption that their polarization states are transmitted unchanged \cite{Ditc}.
By ``circular double refraction'' at the surface of the chiral medium we mean \cite{Hu-L} the phenomenon that the refracted wave splits into two orthogonal circularly polarized components propagating in different directions. It arises from the aforementioned opposite phase shifts due to the phase matching on the boundary.

As noted above, our proof of the nonexistence of circular birefringence is basically a logical reasoning process. It does not depend on the concrete constitutive relations of the chiral medium though the Condon model is used in our discussions. As a matter of fact, it is even unnecessary to employ constitutive relations to do so. 
To demonstrate this, here we revisit the optical rotation in helically-coiled optical fibers. As is known, such an optical rotation is not expressed by constitutive relations. It is a purely geometrical effect, arising from the geometry of the path followed by the fiber \cite{Papp-H, Ross, Varn-BP, Tomi-C, Qian-H, Birc}. But akin to the optical rotation expressed by constitutive relations, it is also conventionally interpreted as the circular birefringence \cite{Papp-H, Ross, Varn-BP, Birc, Qian-H, Qian, Napi-U}. 
We will show that like chiral media, helically-coiled optical fibers do not exhibit the circular birefringence, either. Gratifyingly, discussions of this topic also allow to have a clearer look at why the description of the state of polarization in terms of the Jones vector \cite{Jone} is incomplete, a conclusion that we drew previously in Ref. \cite{Hu-L}.

To be honest, the optical rotation in helically-coiled fibers is distinguished from that expressed by constitutive relations. About half a century ago, Papp and Harms \cite{Papp-H} experimentally observed that ``the plane of polarization is not really rotated along the fiber, and the optical activity exists only with respect to the coordinate system along the fiber.'' 
The first half of the conclusion, ``the plane of polarization is not really rotated along the fiber,'' implies that the rotation observed by Papp and Harms is not the property of the polarization vector in the laboratory reference frame. It actually rules out the possibility of the existence of circular birefringence in the helically-coiled fiber. Regretfully, so important a finding has not received the attention it deserves. 
As a matter of fact, the quantity that Papp and Harms used to describe the observed rotation is the Jones vector. We will argue from the second half of their conclusion, ``the optical activity exists only with respect to the coordinate system along the fiber,'' that the Jones vector is mathematically an entity in the reference frame associated with the local propagation direction. It can only describe the state of polarization relative to the local reference frame. This explains why the Jones vector is not able to completely describe the state of polarization of a plane wave. 
On this basis, we will further show that what Papp and Harms observed is the rotation of the Jones vector in the Serret-Frenet frame \cite{Coxe} or, more precisely, the rotation of the Tang frame \cite{Tang} relative to the Serret-Frenet frame. 
The contents are arranged as follows.

For the sake of convenience, Section \ref{Review} gives a review of the conventional interpretation of the optical rotation in the helically-coiled fiber. Eqs. (\ref{Ar+Al}) were interpreted as expressing the circular birefringence, which was considered to be responsible for the optical rotation determined by Eq. (\ref{DE-AFS}) for the Jones vector $\tilde A$.
The nonexistence of circular birefringence is proven in Section \ref{Nonexistence} through logical reasoning similar to that in Refs. \cite{Hu-L, Hu-L24}. The problem with the conventional interpretation is also analyzed. 
To figure out the meaning conveyed by Papp and Harms' statement that ``the plane of polarization is not really rotated along the fiber,'' a different Jones vector $\tilde{A}'$ that remains unchanged down the fiber is found out. This is achieved in Section \ref{NJV} by making use of the Tang frame, a local reference frame that does not rotate about the instantaneous direction of the fiber.
It is argued in Section \ref{PMofJV}, relying on Papp and Harms' experimental result, that the Jones vectors $\tilde A$ and $\tilde{A}'$ are mathematically entities in the Serret-Frenet and Tang frames, respectively. They only describe the state of polarization relative to the respective local reference frames. On this basis, it is further demonstrated in Section \ref{Reinterpret} that the rotation observed by Papp and Harms just reflects the rotation of the Tang frame relative to the Serret-Frenet frame. Explicit transformations from the polarization vector $\mathbf A$ in the laboratory reference frame into Jones vectors $\tilde A$ and $\tilde{A}'$ are also put forward.
Section \ref{Conclusions} concludes the paper with remarks. It is particularly pointed out that one cannot have a Jones vector to describe the state of polarization relative to the laboratory reference frame.

\section{\label{Review} Review of conventional interpretation}

\begin{figure}[h]
	\centerline{\includegraphics[width=7.0cm]{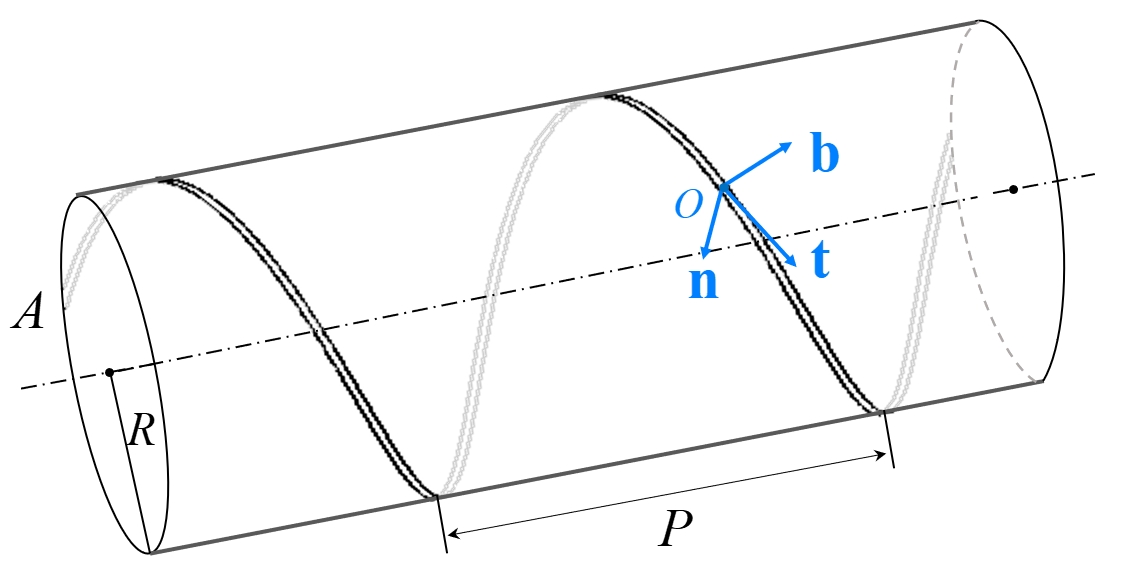}}
	\caption{\label{GoF} Helically-coiled optical fiber showing the Serret-Frenet frame $\mathbf{tnb}$ at a point $O$.}
\end{figure}
The axis of a helically-coiled optical fiber is described mathematically as a twisted curve with a constant torsion $\tau$ and a constant curvature $\chi$ \cite{Coxe}.
The geometry of the fiber is schematically shown in Fig. \ref{GoF}, where $P$ is the pitch and $R$ is the radius of the helix. The helically-coiled fiber is a periodic structure. The arc length per turn is $S=2 \pi/T$, where 
$T=(\chi^2+\tau^2)^{1/2}$. The radius of the fiber core is assumed to be much smaller than $R$ so that the linear birefringence induced by bending the fiber is negligibly small \cite{Ulri-RE}. 
The Serret-Frenet frame at a point $O$ on the fiber axis is an orthogona1 trihedron of unit vectors denoted by $\mathbf t$, $\mathbf n$, and $\mathbf b$, which obey
\begin{equation}\label{ortho}
	|\mathbf{t}|=|\mathbf{n}|=|\mathbf{b}|=1, \quad
	\mathbf{t} =\mathbf{n} \times \mathbf{b},
\end{equation}
where $\mathbf t$ is the tangent representing the instantaneous direction of the fiber, $\mathbf n$ is the principal normal, and $\mathbf b$ is the binormal. They satisfy the Serret-Frenet formulae,
\begin{subequations}\label{FSEs}
\begin{align}
	\frac{d \mathbf{t}}{ds} &=  \chi \mathbf{n},                  \label{DEfort} \\
	\frac{d \mathbf{n}}{ds} &= -\chi \mathbf{t} +\tau \mathbf{b}, \label{DEforn} \\
	\frac{d \mathbf{b}}{ds} &= -\tau \mathbf{n},                  \label{DEforb}
\end{align}
\end{subequations}	
where $s$ is the directed arc measured along the fiber from a fixed point $A$ to the variable point $O$.

Consider a monochromatic wave traveling down the fiber. In the plane-wave approximation, its electric field at a point of distance $s$ can be written as \cite{Papp-H, Ross, Fang-L}
\begin{equation}\label{EF}
	\mathbf{E}(s)=\mathbf{A}(s) \exp(iks),
\end{equation}
where the time dependence $\exp(-i\omega t)$ is assumed and suppressed, $\omega$ is the angular frequency, the unit vector $\mathbf A$ denotes the state of polarization, $k=n k_0$, $n$ is the refraction index of the fiber core, and $k_0$ is the wavenumber in free space. Because only the polarization property of the wave (\ref{EF}) is addressed, its amplitude has been normalized.  
As the polarization vector, $\mathbf A$ is transverse to the local propagation direction, the tangent $\mathbf t$, 
\begin{equation}\label{TC}
	\mathbf{t} \cdot \mathbf{A}=0,
\end{equation}
and therefore can be mathematically expanded as
\begin{equation}\label{AinFSF}
	\mathbf{A}(s)=A_n (s) \mathbf{n} +A_b (s) \mathbf{b},
\end{equation}
where $A_n$ and $A_b$ are the projections onto $\mathbf n$ and $\mathbf b$, respectively, satisfying
$|A_n|^2 +|A_b|^2 =1$. 
To simplify the analysis, we assume that no linear birefringence exists in the fiber. In this case, $A_n$ and $A_b$ obey the following coupled equations \cite{Ross, Qian-H, Fang-L, Frin-D},
\begin{equation}\label{CEs}
		\frac{d A_n}{ds} = \tau A_b,  \quad
		\frac{d A_b}{ds} =-\tau A_n.
\end{equation}
Letting
$\tilde{A}(s)=\bigg(\begin{array}{c}
 	                    A_n\\A_b
                    \end{array}\bigg)$,
which is the well-known Jones vector, the coupled equations can be converted into \cite{Ross, Frin-D, Berr}
\begin{equation}\label{DE-AFS}
	\frac{d \tilde{A}}{ds}
   =\bigg(\begin{array}{cc}
   	           0    & \tau \\
   	          -\tau & 0
          \end{array}\bigg) \tilde{A}.
\end{equation}
It has solutions of right-handed and left-handed circularly polarized modes, given by
\begin{equation}\label{Ar+Al}
	\tilde{A}_r (s)=\frac{\exp(i \tau s)}{\sqrt 2} 
	                \bigg(\begin{array}{c}
		                      1 \\ i
	                      \end{array} \bigg),      \quad
    \tilde{A}_l (s)=\frac{\exp(-i \tau s)}{\sqrt 2} 
                    \bigg(\begin{array}{c}
    	                      1 \\ -i
                          \end{array} \bigg),                   
\end{equation}
respectively. 
Eq. (\ref{DE-AFS}) determines the rotation of $\tilde A$ with a per length rate $-\tau$. It is conventionally interpreted \cite{Ross} as expressing the rotation of the polarization state along the fiber. Accordingly, solutions (\ref{Ar+Al}) are interpreted as expressing the circular birefringence, with $k+\tau$ and $k-\tau$ being interpreted \cite{Qian-H} as the propagation constants of the right-handed and left-handed circularly polarized modes, respectively.

Let us prove that the helically-coiled optical fiber does not exhibit the circular birefringence and explain what is wrong with the above interpretation.

\section{\label{Nonexistence} No circular birefringence exists}

It is well known that the following superposition of the two circularly polarized modes (\ref{Ar+Al}) is linearly polarized, 
\begin{equation*}
	\tilde{A}_1 (s) =\frac{1}{\sqrt 2} (\tilde{A}_r +\tilde{A}_l)
	                =\bigg(\begin{array}{c}
	                	       \cos \tau s \\
	                	      -\sin \tau s 
	                       \end{array}\bigg).
\end{equation*}
Meanwhile, a superposition of the form
\begin{equation*}
	\tilde{A}_2 (s) =\frac{-i}{\sqrt 2} (\tilde{A}_r -\tilde{A}_l)
	                =\bigg(\begin{array}{c}
		                       \sin \tau s \\
		                       \cos \tau s 
	                       \end{array}\bigg)
\end{equation*}
is also linearly polarized. Needless to say, these two linearly polarized modes are rotated in the same way and can be re-expressed as
\begin{equation}\label{A1+A2}
	\tilde{A}_1 (s) =R(s) \bigg(\begin{array}{c}
	                 	            1 \\
	                 	            0 
	                 \end{array}\bigg),            \quad
	\tilde{A}_2 (s) =R(s) \bigg(\begin{array}{c}
	                                0 \\
	                                1 
                           \end{array}\bigg),             
\end{equation}
where
\begin{equation}\label{RM}
R(s)=\bigg(\begin{array}{cc}
	           \cos \tau s & \sin \tau s \\
	          -\sin \tau s & \cos \tau s
           \end{array} \bigg)
\end{equation}
is the matrix for a rotation through an angle $-\tau s$ \cite{Ross}. Because no linear birefringence is assumed to exist in the fiber, they should propagate at the same velocity.
More importantly, they are orthogonal to each other. Taking them as basis states, one can expand the two circularly polarized modes (\ref{Ar+Al}) in the following way,
\begin{equation*}
	\tilde{A}_r (s)=\frac{1}{\sqrt 2} (\tilde{A}_1 +i\tilde{A}_2), \quad
	\tilde{A}_l (s)=\frac{1}{\sqrt 2} (\tilde{A}_1 -i\tilde{A}_2).
\end{equation*}
As a result, whether the right-handed circularly polarized or the left-handed circularly polarized mode propagates at the velocity of the linearly polarized modes (\ref{A1+A2}). That is to say, the helically-coiled optical fiber does not exhibit the circular birefringence.

Let us look at this result from a slightly different point of view. Eq. (\ref{DE-AFS}) has solution of the form \cite{Papp-H,Ross},
\begin{equation}\label{RofAFS}
	\tilde{A} (s)=R(s) \tilde{\alpha},
\end{equation}
where 
$\tilde{\alpha} =\bigg(\begin{array}{c}
	                       \alpha_1 \\ \alpha_2
                       \end{array} \bigg)$
is the ``initial value'' satisfying the normalization condition
$\tilde{\alpha}^\dag \tilde{\alpha}=1$. 
Generally speaking, it is elliptically polarized. Considering that
$R(s)=\exp(i \hat{\sigma}_3 \tau s)$ where
$\hat{\sigma}_3 =\bigg(\begin{array}{cc}
	                       0 & -i \\
	                       i &  0
                       \end{array}\bigg)$,
its polarization ellipticity, given by 
\begin{equation*}
	\sigma =\tilde{A}^\dag \hat{\sigma}_3 \tilde{A}
	=\tilde{\alpha}^\dag \hat{\sigma}_3 \tilde{\alpha},
\end{equation*}
does not change along the fiber.
It is noted that solution (\ref{RofAFS}) can be expanded in terms of the orthogonal linearly polarized modes (\ref{A1+A2}) as
\begin{equation*}
	\tilde{A}(s) =\alpha_1 \tilde{A}_1 (s) +\alpha_2 \tilde{A}_2 (s).
\end{equation*}
Now that $\tilde{A}_1 (s)$ and $\tilde{A}_2 (s)$ have the same propagation velocity, the elliptically polarized mode (\ref{RofAFS}) should also propagate at this velocity no matter what its polarization ellipticity is. Therefore, the two circularly polarized modes (\ref{Ar+Al}) cannot propagate at different velocities. 

Then, what is wrong with the conventional interpretation? To answer this question, we point out that the state of polarization described by the Jones vector (\ref{RofAFS}) is always rotated along the fiber regardless of its polarization ellipticity. Such a rotation shows up as the rotation of the orientation of polarization ellipse \cite{Ross}. As a corollary, the circularly polarized modes (\ref{Ar+Al}), as special cases of elliptically polarized mode, should also be thought of as undergoing the same rotation. In fact, from Eq. (\ref{RofAFS}) it follows that solutions (\ref{Ar+Al}) can be re-expressed as 
\begin{equation*}
	\tilde{A}_r (s)=R(s) \frac{1}{\sqrt 2} 
	\bigg(\begin{array}{c}
		      1 \\ i
	      \end{array} \bigg),             \quad
	\tilde{A}_l (s)=R(s) \frac{1}{\sqrt 2} 
	\bigg(\begin{array}{c}
		      1 \\ -i
	      \end{array} \bigg),                   
\end{equation*}
because
$\frac{1}{\sqrt 2} \bigg(\begin{array}{c}
	                         1 \\ \pm i
                         \end{array} \bigg)$
is the eigenvector of $\hat{\sigma}_3$ with eigenvalue $\pm 1$.
This indicates that the states of polarization of the two circularly polarized modes are rotated along the fiber in the same way. Solutions (\ref{Ar+Al}) just mean that such rotations give rise to opposite phase shifts. 
These phase shifts are mistakenly considered in conventional interpretation as arising from their different propagation velocities under the implicit assumption that their polarization states are not rotated along the fiber \cite{Russ-BW}.

In addition, according to expression (\ref{EF}), in order to make that interpretation, one also needs to implicitly assume \cite{Hecht, Gold} that the polarization vector $\mathbf{A} (s)$ can be fully replaced with the Jones vector $\tilde{A} (s)$. 
After such a replacement, expression (\ref{EF}) becomes \cite{Ross}
\begin{equation*}
	\mathbf{E}(s)= \tilde{A}(s) \exp(iks).
\end{equation*}
It is seen that without this assumption, one would not be able to deduce from solutions (\ref{Ar+Al}) that the right-handed and left-handed circularly polarized waves have propagation constants $k+\tau$ and $k-\tau$, respectively.
However, with the help of Eqs. (\ref{DEforn}), (\ref{DEforb}), and (\ref{CEs}), one finds 
\begin{equation}\label{DE-EFV}
	\frac{d\mathbf{A}}{ds} =-\chi  A_n \mathbf{t}.
\end{equation}
It is different from Eq. (\ref{DE-AFS}). It is even not an equation about a rotation. To replace the polarization vector $\mathbf A$ with the Jones vector $\tilde A$ is totally wrong. We thus again arrive at the conclusion \cite{Hu-L} that the Jones vector is not able to completely describe the state of polarization denoted by the polarization vector.

Heretofore, we have shown mathematically that in contrast with the optical rotation in chiral media, the rotation mentioned by Papp and Harms in their conclusion ``the optical activity exists only with respect to the coordinate system along the fiber'' is not the property of the polarization vector satisfying Eq. (\ref{DE-EFV}). Instead, as Papp and Harms analyzed, it is the property of the Jones vector expressed by Eq. (\ref{RofAFS}). This is shown by the fact that Eq. (7) in Ref. \cite{Papp-H} reduces to Eq. (\ref{RofAFS}) here when the linear birefringence expressed by Eq. (6) in Ref. \cite{Papp-H} is ignored.
Nevertheless, Eq. (\ref{DE-EFV}) does not mean that the state of polarization denoted by $\mathbf A$ remains unchanged down the fiber as is conveyed by Papp and Harms' statement that ``the plane of polarization is not really rotated along the fiber.'' 
To fully understand Papp and Harms' experimental result, it is beneficial to make clear what the statement means exactly.

\section{\label{NJV}Introduction of a new Jones vector}

As is known, the Serret-Frenet frame is a local reference frame that rotates about the tangent $\mathbf t$ with a per length rate $\tau$. 
If, according to Papp and Harms, the optical rotation exists only with respect to the Serret-Frenet frame, we consider a local reference frame with zero rate of rotation about $\mathbf t$. Obviously, the result of reversely rotating the Serret-Frenet frame about $\mathbf t$ with a per length rate $-\tau$ will meet the need.
Let the unit vectors for the transverse axes of such a frame be $\mathbf u$ and $\mathbf v$, which are given by
\begin{equation}\label{RofTF}
	\mathbf{u}=\exp[i(\mathbf{\Sigma} \cdot \mathbf{t})\tau s] \mathbf{n}, \quad
	\mathbf{v}=\exp[i(\mathbf{\Sigma} \cdot \mathbf{t})\tau s] \mathbf{b},
\end{equation}
respectively, where 
$(\Sigma_k)_{ij} =-i \epsilon_{ijk}$ with $\epsilon_{ijk}$ the Levi-Civit\'{a} pseudotensor. 
With the help of Eqs. (\ref{ortho}) and the formula \cite{Norm}
\begin{equation*}
	\exp[i(\mathbf{\Sigma} \cdot \mathbf{x}) \phi] \mathbf{y}
	=\mathbf{y} \cos \phi -\mathbf{x} \times \mathbf{y} \sin \phi 
	+\mathbf{x} (\mathbf{x} \cdot \mathbf{y}) (1-\cos \phi),
\end{equation*} 
where $\mathbf x$ stands for a rotation axis and $\mathbf y$ is any vector, we have
\begin{equation}\label{TofF}
	\mathbf{u}=\mathbf{n} \cos\tau s -\mathbf{b} \sin\tau s, \quad
	\mathbf{v}=\mathbf{n} \sin\tau s +\mathbf{b} \cos\tau s.
\end{equation}
It is seen that so introduced reference frame is one particular Tang frame \cite{Tang, Fang-L}. 
It should be emphasized that the Tang frame shares with the Serret-Frenet frame the same longitudinal axis $\mathbf t$, the local propagation direction of the wave. It differs from the Serret-Frenet frame only in the transverse axes.
The variations of $\mathbf u$ and $\mathbf v$ along the fiber satisfy
\begin{equation}\label{DE-uv}
	\frac{d \mathbf{u}}{ds}=-\chi \mathbf{t} \cos\tau s, \quad
	\frac{d \mathbf{v}}{ds}=-\chi \mathbf{t} \sin\tau s,
\end{equation}
by virtue of Eqs. (\ref{DEforn}) and (\ref{DEforb}).
The inverse transformation of Eqs. (\ref{TofF}) is given by
\begin{equation}\label{ITofF}
	\mathbf{n}= \mathbf{u} \cos\tau s +\mathbf{v} \sin\tau s, \quad
	\mathbf{b}=-\mathbf{u} \sin\tau s +\mathbf{v} \cos\tau s.
\end{equation}

In the Tang frame, the polarization vector of the wave (\ref{EF}) can be expanded as
\begin{equation}\label{AinTF}
	\mathbf{A}(s)=A_u \mathbf{u}(s) +A_v \mathbf{v}(s),
\end{equation}
where $A_u$ and $A_v$ are the projections onto $\mathbf u$ and $\mathbf v$, respectively.
Like $A_n$ and $A_b$ in expansion (\ref{AinFSF}), $A_u$ and $A_v$ here also make up a Jones vector, denoted by
$\tilde{A}' =\bigg(\begin{array}{c}
	A_u\\A_v
\end{array}
\bigg)$.
Substituting Eqs. (\ref{ITofF}) into expression (\ref{AinFSF}) and comparing the result with expression (\ref{AinTF}), we have
\begin{equation}\label{Au+Av}
	\begin{aligned}
		A_u &= A_n \cos\tau s -A_b \sin\tau s, \\
		A_v &= A_n \sin\tau s +A_b \cos\tau s,
	\end{aligned}
\end{equation}
which can be rewritten in terms of the rotation matrix (\ref{RM}) as
\begin{equation}\label{TofJV}
	\tilde{A}' =R^\dag (s) \tilde{A}.
\end{equation}
With the help of Eq. (\ref{DE-AFS}), it is easy to obtain from Eq. (\ref{TofJV}) that
\begin{equation}\label{DE-AT}
	\frac{d \tilde{A}'}{ds}=0.
\end{equation}
Instead of the polarization vector $\mathbf A$, it is the Jones vector $\tilde{A}'$ that remains unchanged down the fiber! In fact, upon substituting Eq. (\ref{RofAFS}) into Eq. (\ref{TofJV}), we have
\begin{equation}\label{alpha}
	\tilde{A}' =\tilde{\alpha},
\end{equation}
which is independent of $s$.
After having obtained Eqs. (\ref{DE-AFS}) and (\ref{DE-AT}), we are ready to discuss why the Jones vector is not able to completely describe the polarization state of light and what Papp and Harms really observed.

\section{\label{PMofJV} Distinguishing Jones vector from polarization vector}

\subsection{Jones vector as mathematical entity in local reference frame}

As is shown by expressions (\ref{AinFSF}) and (\ref{AinTF}), both the Jones vectors $\tilde A$ and $\tilde{A}'$ are defined for the same polarization vector $\mathbf A$. Nevertheless, whether Eq. (\ref{DE-AFS}) for $\tilde A$ or Eq. (\ref{DE-AT}) for $\tilde{A}'$ is different from Eq. (\ref{DE-EFV}) for $\mathbf A$. 
To explain this discrepancy, we note that when it is written down, expression (\ref{EF}) is assumed to be the solution of Maxwell's equations in the laboratory reference frame. This is to say that the polarization vector $\mathbf A$ is mathematically an entity in the laboratory reference frame. That Eqs. (\ref{DE-AFS}) and (\ref{DE-AT}) are different from Eq. (\ref{DE-EFV}) means that neither of $\tilde A$ and $\tilde{A}'$ can be thought of as an entity in the laboratory reference frame. 
Furthermore, the difference between Eq. (\ref{DE-AFS}) and Eq. (\ref{DE-AT}) means that $\tilde A$ and $\tilde{A}'$ cannot be thought of as entities in a same reference frame. 
Indeed, as is also shown by expressions (\ref{AinFSF}) and (\ref{AinTF}), $\tilde A$ and $\tilde{A}'$ are introduced on different bases. The former is on the basis of the Serret-Frenet frame; the latter is on the basis of the Tang frame. 
Let us make use of Papp and Harms' experimental result to argue that the Jones vectors $\tilde A$ and $\tilde{A}'$ are entities in the Serret-Frenet and Tang frames, respectively.

As mentioned in the last paragraph of Section \ref{Nonexistence}, the rotation observed by Papp and Harms is the property of the Jones vector $\tilde A$ satisfying Eq. (\ref{RofAFS}). They did the experiment with linearly polarized incident light. Their experimental result is expressed in Ref. \cite{Papp-H} by Eq. (5), which is reproduced here as
\begin{equation}\label{ER}
	\Omega/n=-\tau S,
\end{equation} 
where $\tau$ takes the role of the torsion $w$ in Ref. \cite{Papp-H}. 
According to Papp and Harms, the left side is the measured angle of rotation of the polarization plane for one turn of the fiber. Remembering that matrix (\ref{RM}) represents a rotation through an angle $-\tau s$, the right side is exactly the rotation angle of $\tilde A$ predicted by Eq. (\ref{RofAFS}) for the length $S$ per turn of the fiber. 
It is on the basis of Eq. (\ref{ER}) that Papp and Harms concluded ``the optical activity exists only with respect to the coordinate system along the fiber.'' We thus have reason to think of the Jones vector $\tilde A$ as a mathematical entity in the Serret-Frenet frame.
If this is the case, the Jones vector $\tilde{A}'$ is an entity in the Tang frame. In view of this, we can say that it is the Jones vector in the Tang frame that remains unchanged down the fiber.
This should be the first time to identify that the Jones vector and the polarization vector are not quantities in the same reference frame.

\subsection{Jones vector describes state of polarization relative to local reference frame}

It is time for us to explain why the Jones vector cannot completely describe the state of polarization.
Now that the polarization vector $\mathbf A$ is a quantity in the laboratory reference frame, the state of polarization denoted by $\mathbf A$ is a physical phenomenon with respect to the laboratory reference frame.
With this in mind, Eq. (\ref{DE-EFV}) determines the way in which the state of polarization varies down the fiber with respect to the laboratory reference frame.
Of course, transversality condition (\ref{TC}) makes it possible to expand the polarization vector $\mathbf A$ in any local reference frame associated with the propagation direction $\mathbf t$. Expressions (\ref{AinFSF}) and (\ref{AinTF}) are the expansions in the Serret-Frenet and Tang frames, respectively. Upon taking Eq. (\ref{DE-AT}) into consideration, one readily gets from expression (\ref{AinTF}) and Eq. (\ref{DE-uv})
\begin{equation*}
	\frac{d \mathbf{A}}{ds} =-\chi (A_u \cos \tau s +A_v \sin \tau s) \mathbf{t}.
\end{equation*}
It is equivalent to Eq. (\ref{DE-EFV}) because of 
$A_n =A_u \cos \tau s +A_v \sin \tau s$
by virtue of Eqs. (\ref{Au+Av}). 
From these discussions we can say that the Jones vector $\tilde A$ plays the role of describing the state of polarization relative to the Serret-Frenet frame. Accordingly, the Jones vector $\tilde{A}'$ plays the role of describing the state of polarization relative to the Tang frame.
In a word, the Jones vector as a quantity in the local reference frame can only describe the state of polarization relative to that reference frame. 
Mathematically, this is realized, as is implied in expression (\ref{AinFSF}) or (\ref{AinTF}), by transforming the polarization vector $\mathbf A$ in the laboratory reference frame into the Jones vector in the local reference frame.

We have seen that the Jones vector $\tilde{A}'$ in the Tang frame and the Jones vector $\tilde A$ in the Serret-Frenet frame behave differently. 
The former is invariant along the fiber, indicating that the state of polarization relative to the Tang frame is transmitted unchanged. The latter is rotated along the fiber with a rotation rate $-\tau$. This shows that relative to the Serret-Frenet frame, the same state of polarization is rotated along the fiber. 
However, because the Tang frame is also rotated relative to the Serret-Frenet frame with the rotation rate $-\tau$, the invariance of $\tilde{A}'$ in the Tang frame means that the rotation of $\tilde A$ in the Serret-Frenet frame is just the rotation of the Tang frame relative to the Serret-Frenet frame. In other words, the rotation observed by Papp and Harms is the reflection of the rotation of the Tang frame relative to the Serret-Frenet frame.
Let us show this in more detail below.

\section{\label{Reinterpret} Reinterpretation of Papp and Harms's observation}

As mentioned above, expression (\ref{AinFSF}) implies that the polarization vector $\mathbf A$ in the laboratory reference frame is mathematically transformed into the Jones vector $\tilde A$ in the Serret-Frenet frame.
Our first step is to change it into an equation to explicitly express that $\tilde A$ is a quantity in the Serret-Frenet frame. This is done by introducing the 3-by-2 matrix
$\varpi =(\begin{array}{lr}
	          \mathbf{n} & \mathbf{b}
          \end{array})$
to convert it into 
\begin{equation}\label{ApiFS}
	\tilde{A}= \varpi^\dag \mathbf{A},
\end{equation}
where the unit vectors $\mathbf n$ and $\mathbf b$ are regarded as column matrices of three elements, which are their Cartesian components in the laboratory reference frame, the superscript $\dag$ denotes the conjugate transpose,
and the rule of matrix multiplication is employed.
To show this, it is noted that expression (\ref{AinFSF}) can be rewritten in terms of $\varpi$ as \cite{Li}
\begin{equation}\label{QUT}
	\mathbf{A}=\varpi \tilde{A}.
\end{equation}
Moreover, the matrix $\varpi$ has the property
\begin{equation*}
	\varpi^\dag \varpi =I_2,
\end{equation*}
by virtue of Eqs. (\ref{ortho}), where $I_2$ is the 2-by-2 unit matrix. Multiplying both sides of Eq. (\ref{QUT}) with $\varpi^\dag$ from the left and taking this property into account, we will arrive at Eq. (\ref{ApiFS}).
It is pointed out that the unit vectors that make up the matrix $\varpi$ specify the orientation of the transverse axes of the Serret-Frenet frame with respect to the laboratory reference frame. Because they are always perpendicular to the instantaneous longitudinal axis $\mathbf t$, the matrix $\varpi$ takes the role of denoting the entire Serret-Frenet frame with respect to the laboratory reference frame.
In view of this, Eq. (\ref{ApiFS}) shows that $\tilde A$ is the projection of the polarization vector $\mathbf A$ onto the Serret-Frenet frame denoted by $\varpi$. In other words, Eq. (\ref{ApiFS}) explicitly expresses that the Jones vector $\tilde A$ is a quantity in the Serret-Frenet frame.
Similarly, to explicitly express that the Jones vector $\tilde{A}'$ is a quantity in the Tang frame, we introduce 
$\varpi' =(\begin{array}{lr}
	           \mathbf{u} & \mathbf{v}
           \end{array})$
to denote the Tang frame with respect to the laboratory reference frame, which is connected with $\varpi$ via
\begin{equation}\label{RofTF2}
	\varpi' =\exp[i(\mathbf{\Sigma} \cdot \mathbf{t})\tau s] \varpi,
\end{equation}
by virtue of Eqs. (\ref{RofTF}). Eq. (\ref{RofTF2}) describes the rotation of the Tang frame relative to the Serret-Frenet frame with a rotation rate $-\tau$. Like $\varpi$, $\varpi'$ also has the property
\begin{equation*}
	\varpi'^\dag \varpi'  =I_2.
\end{equation*}
In terms of $\varpi'$, $\tilde{A}'$ is related to the polarization vector $\mathbf A$ by
\begin{equation}\label{ApiT}
	\tilde{A}' = \varpi'^\dag \mathbf{A},
\end{equation}
meaning that $\tilde{A}'$ is the projection of the polarization vector $\mathbf A$ onto the Tang frame denoted by $\varpi'$.

It is worth noting that according to Eqs. (\ref{TofF}), the rotation of $\varpi'$ relative to $\varpi$ can also be expressed as
\begin{equation}\label{Tofpi}
	\varpi' =\varpi R(s),
\end{equation}
where $R(s)$ is given by Eq. (\ref{RM}). The rotation matrix $R(s)$ in this equation should be viewed as acting to the left.
Eq. (\ref{Tofpi}) can be rewritten as 
\begin{equation*}
	\varpi=\varpi' R^\dag (s).
\end{equation*}
Upon substituting it into Eq. (\ref{ApiFS}) and taking Eq. (\ref{ApiT}) into account, we get
\begin{equation*}
	\tilde{A}= R(s) \varpi'^\dag \mathbf{A} =R(s) \tilde{A}',
\end{equation*}
which is Eq. (\ref{RofAFS}) by virtue of Eq. (\ref{alpha}). It is clearly seen from these derivations that the rotation of $\tilde A$ in the Serret-Frenet frame is indeed the rotation of the Tang frame relative to the Serret-Frenet frame. That is to say, what Papp and Harms observed just reflects the rotation of the Tang frame relative to the Serret-Frenet frame.
As a corollary, both the phase factors $\exp(i\tau s)$ and $\exp(-i\tau s)$ in Eqs. (\ref{Ar+Al}) are the reflection of this rotation. They have nothing to do with the propagation velocity of the two orthogonal circularly polarized modes.

\section{Conclusions and Remarks}\label{Conclusions}

In conclusion, we proved by use of logical reasoning that expressions (\ref{Ar+Al}) do not mean the circular birefringence.
Moreover, we argued with the help of Eq. (\ref{ER}), the rotation observed by Papp and Harms, that the Jones vector $\tilde A$ given by Eq. (\ref{ApiFS}) and the Jones vector $\tilde{A}'$ given by Eq. (\ref{ApiT}) are quantities in the Serret-Frenet and Tang frames, respectively. They can only describe the state of polarization relative to their respective local reference frames. Though the former is rotated down the fiber in accordance with Eq. (\ref{DE-AFS}), the latter is invariant as is shown by Eq. (\ref{DE-AT}).
On this basis, we further demonstrated that what Papp and Harms observed is actually the rotation of the Tang frame relative to the Serret-Frenet frame. 

It is emphasized that Eq. (\ref{DE-EFV}) for the variation of the polarization vector $\mathbf A$ along the fiber is different from Eq. (\ref{DE-AFS}) for the rotation of the Jones vector $\tilde A$ in the Serret-Frenet frame. 
But the relationship between $\mathbf A$ and $\tilde A$ is given by Eq. (\ref{AinFSF}).
Since the solution to Eq. (\ref{DE-AFS}) is given by Eq. (\ref{RofAFS}), in order to know how the polarization vector varies along the fiber, it is only required to find the solutions to Eqs. (\ref{FSEs}) under certain initial conditions. A discussion of this issue is beyond the scope of the present paper.

It is noted, as is expressed by Eqs. (\ref{ApiFS}) and (\ref{ApiT}), that the local reference frame in which the Jones vector is defined is associated with the propagation direction.
Because the propagation direction of the wave traveling down a helically-coiled fiber is not fixed, the laboratory reference frame cannot serve as one of local reference frames to define its Jones vector.
That is to say, one cannot have a Jones vector that describes the state of polarization relative to the laboratory reference frame. 

In a word, we showed, through reexamining the optical rotation in a helically-coiled fiber, that the Jones vector as a quantity in the local reference frame is not able to completely describe the polarization state of a plane wave.
In order to make use of the Jones vector to do so, the local reference frame in which the Jones vector is defined must be specified simultaneously. 
It is hoped that the findings presented here will deepen understanding of the phenomenon of optical polarization.



\section*{Acknowledgments}

This work was supported in part by National Natural Science Foundation of China under Grant No. 62075134.

\section*{Disclosures}
The authors declare no conflicts of interest.

\end{document}